\documentclass[twocolumn]{article}

\usepackage{amsmath}
\usepackage{amssymb}
\usepackage{graphicx}

\newcommand{\helptext}[1]{}
\newcommand{\marginhelp}[1]{}

\newcommand{\be}{\begin{equation}}
\newcommand{\ee}{\end{equation}}

\newcommand{\de}{\stackrel{\mbox{\tiny def}}{=}}

\newcommand{\rmd}{{\mathrm d}}

\newcommand{\prob}[2]{P(#1|#2)}
\newcommand{\probdist}[2]{p(#1|#2)}

\newcommand{\proposition}[2]{#1_{#2}}

\begin{document}

\title{Measuring Portfolio Diversification}
\author{Ulrich Kirchner \\[0.25cm]
\small ICAP \\ \small PO Box 1210, Houghton, 2041, South Africa \\ \small ulrich.kirchner@icap.co.za\\[1cm]
Caroline Zunckel\\[0.25cm]
\small Astrophysics and Cosmology Research Unit, University of Kwazulu-Natal\\
\small Westville, Durban, 4000, South Africa\\[1cm]
}

\maketitle

\begin{abstract}
In the market place, diversification reduces risk and provides protection against extreme events by ensuring that
one is not overly exposed to individual occurrences.  We argue that diversification is best
measured by characteristics of the combined portfolio of assets and
introduce a measure based on the information entropy of the probability distribution for the final portfolio asset value.
For Gaussian assets the measure is a logarithmic function of the variance and combining independent Gaussian assets of equal variance
{\em adds} an amount to the diversification.

The advantages of this measure include that it naturally extends to any type of distribution and that it takes all moments into account.
Furthermore, it can be used in cases of undefined weights (zero-cost assets) or moments.
We present examples which apply this measure to derivative overlays.
\end{abstract}

\section{Introduction}
Combining different assets in a portfolio changes the return and risk\footnote{We will leave the precise definition of ``risk'' here open.} characteristics.
Diversification strategies allow  to fine tune for risk appetite and parameter ranges, which are usually stipulated in investment mandates.

While diversification is an intuitive concept, there is no unique quantitative measure of it.
The benefits of portfolio diversification clearly stem from the independence or the offsetting of asset value changes. 
Hence any good measure should take asset characteristics
into account in addition to portfolio exposures to each asset
(e.g., asset weights in the portfolio).

Here we want to give a brief review of existing measures before introducing a new measure based on the entropy of the
return probability distribution (for the total portfolio). This measure naturally extends to zero-cost assets (for which weights cannot be defined)
and we show the diversification benefits of some derivative overlays.

\section{Existing Measures}

Measures have been suggested (for an overview see for example \cite{woerheide}), which fall broadly into two categories: just weight based and taking return characteristics into account.
\begin{description}
\item[Herfindahl Index]
A simple and widely used weight-based diversification index given by
\be
1-\sum_{i=1}^N w_i^2,
\ee
where the $w_i$ are the weights for the $N$ assets. This measure does not account for
inter-dependence between assets and differing risk characteristics.

\item[Weight Entropy]
One way to interpret portfolio weights is to see them as the probability of a
``randomly''\footnote{Here in the sense of equal likelihood. It might indeed be impossible to define randomness consistently.}
choosen currency unit to be invested in a certain asset. One could then argue that the entropy difference between these probabilities and the uniform distribution is a measure of information content
and diversification. The corresponding measure is the weight entropy
\be
- \sum_{i=1}^N w_i \log(w_i).
\ee
This measure also has an intriguing sub-division property, which relates the overall entropy to the entropy of sub-portfolios and the weights of the sub-portfolios.
\be
E=-w_i \log(w_i) + \sum w_i E_i,
\ee
where $w_i$ are the portfolio weights and $E_i$ are the entropies of the sub-portfolios.

\item[Portfolio Diversification Index]
This measure tries to assess how many independent bets there are using the eigenvalues of the covariance matrix
of the returns of individual assets making up a combined portfolio (which is usually estimated from
historic data). It is given by
\be
2 \sum_{k=1}^M k \lambda_k -1,
\ee
where $\lambda_k$ are the ordered and normalized covariance eigenvalues.
In its original form it does not deal with the actual weights assigned to the assets. However, they can be incorporated by considering weighted returns, i.e.,
using the covariance matrix of the fractional contributions of the assets to the total portfolio performance.

An important shortcoming of this measure is its sole reliance on the covariance matrix, which  is the second moment of a multi-variate distribution.
Only for the special case of (multi-variate) normality does this take all distributional information into account (besides the expected value).  
\end{description}

\section{Interpretations and Notation}
We follow the Bayesian interpretation of probability \cite{jaynes} as a measure of degree of believe, which is always dependent on subjective
background information.
The probability of proposition $A$ to be true, given that $B$ is true is denoted by $\prob{A}{B}$. $A$ and $B$ can be composed of several
propositions. $I$ denotes the available (subjective) background information.

Similarly, probability distributions are denoted by $\probdist{\proposition{A}{x}}{B}$, where $\proposition{A}{x}$ is a proposition involving a continuous variable $x$.

Here our primary interest is in probability distributions for the prices of financial assets. Such distributions can be based on subjective beliefs, or they can be
market-implied distributions, which can be extracted from traded derivative prices \cite{uk-pa,uk-mip}.

\section{A Measure based on Entropy}
Information entropy is known as a measure of dispersion in a distribution. In contrast to the variance, it measures the expected information gained per outcome.\footnote{
For a discrete probability distribution the information entropy is invariant under the exchange of any two probability values, i.e., it does not depend on the outcome variable,
just on the probabilities of the possible outcomes.} 
We want to argue here that 
the information entropy of the probability distribution of the final value of the portfolio is a natural measure of
diversification. Let us define such a diversification measure $D(A|I)$ as
\be
D(A|I) \de \int_{-\infty}^\infty \probdist{A_a}{I} \log(\probdist{A_a}{I}) \; \rmd a,
\ee
which is just the negative of the information entropy.
Here $\probdist{A_a}{I}$ is the probability distribution for the final value $a$ of portfolio $A$.
Note that the unit depends on the base of the logarithm used --- for the natural logarithm
the diversification will be measured in ``nats'', which differ by a factor $1/\log(2)$ from ``bits''.

As the probability distributions are subjective, i.e., dependent on some subjective background information $I$, so will be the diversification measure.
This means that our judgment of diversification might change due to new {\em information} available.

One pleasing property of this measure is that it does not make any distributional assumptions, but can be applied to any type of
distribution. Obviously, it is in itself a challange to find a suitable probability distribution expressing information and beliefs about the
portfolio returns.

One consequence of the above is that the measure is not just a function of a finite number of moments/cumulants, like the variance.
Instead $D(A|I)$ depends on all moments and remains defined even if moments become undefined (like for power-law distributions). For example, for the Cauchy distribution
\be
\probdist{A_x}{I} = \frac{1}{\pi \gamma} \frac{1}{1+\left( \frac{x-x_0}{\gamma} \right)^2}
\ee
all moments are undefined, but our diversification measure remains well-defined and takes the value
\be
D(A|I) = - \log( 4 \pi \gamma ).
\ee

For the simple case of normally distributed final portfolio values with variance $\sigma^2$ one finds
\be
D(A|I) = -\frac 1 2 \log(2 \pi e \sigma^2 ),
\ee
where $A$ stands for the set of propositions $A_a$ that the final portfolio value will be $a$.

The more peaked the probability distribution, the larger the value for the diversification measure. In the extreme limiting case of a delta function the measure approaches infinity.
Hence one could argue that cash, which generally has the least uncertainty about its future value in currency terms, has the highest diversification of all investable assets.

\section{Motivation}
\begin{figure}[htb]
\begin{center}
\includegraphics[width=2.8in]{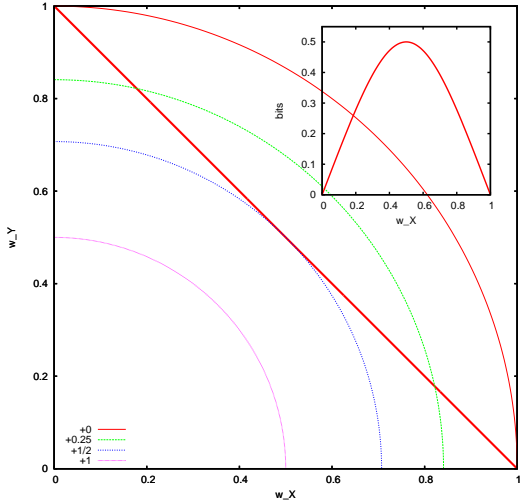}
\end{center}
\caption{Diversification benefit for two uncorrelated assets of equal variance. The change in diversification
depends on the distance to the origin in the weight space. Each circle corresponds to a fixed change in the diversification measure (in bits) and the portfolio weights determine its position along the straight line.
The inset shows the change in diversification measure as a function of $w_X$.}
\label{fig-two-asset-equ-var-div}
\end{figure}

Let us consider the case of two independent assets $A$ and $B$, whose future value is
normally distributed with equal variance and expected value.
Our diversification measure for each asset takes the value
\be
D(A|I) = D(B|I) = -\frac 1 2 \log(2 \pi e \sigma^2 ).
\ee
Let us now form a combined portfolio $C$ by
investing the fraction $w_A$ and $w_B$ of the total portfolio value in asset $A$ and $B$, respectively.
The final value of portfolio $C$ is then given by
\be
c=w_A a+ w_B b.
\ee
As $A$ and $B$ are normal, $C$ is normal with variance
\be
(w_A^2 +w_B^2) \sigma^2
\ee
and the diversification measure for $C$ takes the form
\be
D(C|I)
=D(A|I) - \log\left(\sqrt{w_A^2+w_B^2}\right).
\ee
Hence diversification depends logarithmically on the distance to the origin in weight space.
This is illustrated in Figure \ref{fig-two-asset-equ-var-div}.

This is a sensible behaviour for a diversification measure: for equal variances it is maximized for equal weights (in which case it increases by half a bit)
and the increase due to diversification is independent of asset variances. Furthermore,
it is additive - if the total diversification is $k$ bits, then we can think of it as being the equally weighted mixture of
\be
n=2(k-j)
\ee
independent assets (of equal variancence) each with diversification $j<k$. 

This result can easily be generalized to the case of differing variance $\sigma^2_A$ and $\sigma^2_B$ for $A$ and $B$, respectively.
Using $\sigma_C^2=w_A^2 \sigma_A^2 + w_B^2 \sigma_B^2$ one finds
the result
\begin{eqnarray}
D(C|I) &=& \frac{D(A|I)+D(B|I)}{2} \nonumber \\&& - \log\left(\sqrt{w_A^2\frac{\sigma_A}{\sigma_B}+w_B^2\frac{\sigma_B}{\sigma_A}}\right).
\end{eqnarray}
Here the first term is the (equally weighted) average diversification and the second term is the diversification benefit dependent on the weights and variances.

\section{Transformation Properties}
Let the value $c$ of portfolio $C$ be a strong monotonic increasing function of the value $a$ of portfolio $A$, the ``underlying''.
The portfolio value probability distributions are related by
\be
\probdist{C_{c(a)}}{I} = \probdist{A_a}{I} \left( \frac{\rmd c}{\rmd a} \right)^{-1}
\ee
Substituting this into the definition for the diversification measure one finds
\be
D(C|I) = D(A|I) - \int_{-\infty}^\infty \rmd a \; \probdist{A_a}{I} \log \left( \Delta \right),
\label{eq-tp}
\ee
where
\be
\Delta=\frac{\rmd c}{ \rmd a}
\ee
is the ``future delta'' (the delta at the time for which the diversification measure is evaluated).

One special case of this is constant gearing
\be
c=\lambda a,
\ee
where $\lambda>0$ and the diversification benefit is
\be
D(C|I)=D(A|I)-\log(\lambda).
\ee
This illustrates that gearing ($\lambda>0$), which steepens the dependence of the final portfolio value on the final asset value, decreases diversification.

The results above can be generallized to strongly monotonic decreasing relationships.
If there is no strong monotonic relationship then the domain can be broken into distinct segments, which are either strong monotonic or constant.

\section{Examples}
For the examples below we assume that asset returns are log-normal with constant (log-return) variance $\sigma^2$ and expected growth factor
equal to the risk free rate growth $e^{rt}$. Hence the Black-Scholes formulas can be used for the option pricing.

\subsection{Long dated put}
\begin{figure}[htb]
\begin{center}
\includegraphics[width=2.8in]{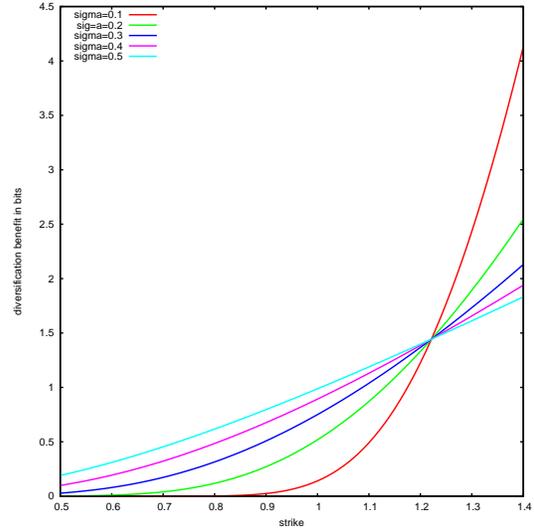}
\end{center}
\caption{Diversification benefit (considered for one year) of a long put position with a two year term. The increase in the diversification
measure is shown for different strikes and volatilities. Here we assumed a risk-free rate and expected value of the log-normal distribution of 10\% per annum.}
\label{fig-db-put}
\end{figure}

We consider here the diversification over a one year
horizon. We augment the asset with a bought put option with expiry in two years and strike price $E$. Let $\tau=1\mathrm{yr}$ be the
remaining life time of the option at the time for which we consider the diversification measure.
The change of the combined portfolio value with the underlying asset is then the sum of future ``deltas''
\be
\Delta =  e^{-D\tau} [N(d_1) -1]+1,
\ee
where $D$ is the (continuous) dividend yield and
\be
d_1 = \frac{\ln(a/E) + (r-D+\sigma^2/2)\tau}{\sigma \sqrt{\tau}}.
\ee
Let us assume here that there is no dividend yield on the asset, i.e., $D=0$.
The diversification benefit is then
\be
- \int_{-\infty}^\infty \rmd a \; \probdist{A_a}{I} \log \left( N(d_1(a)) \right)
\ee
We evaluate this for different strike prices $E$. Figure \ref{fig-db-put} shows the resulting diversification benefits for different standard deviations (for the annual log returns).  

For a given fixed variance, if the strike price of the put option is low, then the final asset value distribution is hardly affected and the diversification effect diminishes.
As the strike price increases more ``negative'' outcomes (asset value below strike price) are eliminated as the put-payoff compensates for the loss. This is equivalent to a narrowing of the
probability distribution and an increase of the diversification measure.
We note that the diversification benefit gained with increasing strike price is far more substantial for less volatile assets.

Note that the option considered here still has one year left to expiry at the time for which we consider the diversification measure.
The above still holds, with the trend becoming more extreme as there is less time to expiry remaining.

\subsection{Put-Spread}

The put spread, being the combination of a bought upper put and a sold lower put, is now considered. 
The totla portfolio future delta is now
\be
\Delta =  e^{-D\tau} [N(d_1(x,E_u)) -N(d_1(x,E_l))]+1,
\ee
where $E_u$ is the strike of the upper put and $E_l$ is the strike of the lower put. 

\begin{figure}[htb]
\begin{center}
\includegraphics[width=2.8in]{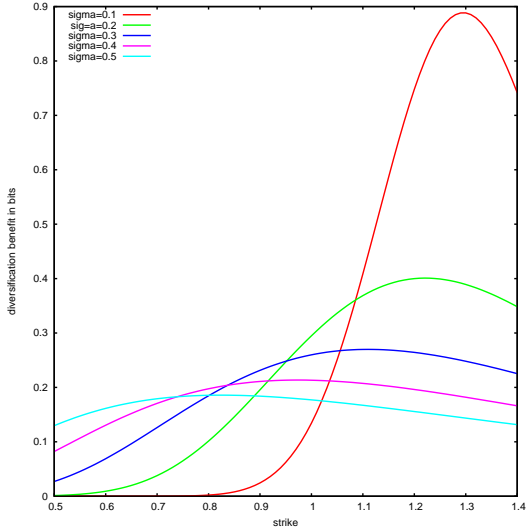}
\end{center}
\caption{Diversification benefit (considered for one year) of a 10\% put-spread position with a two year term. The increase in the diversification
measure is shown for different strikes of the upper put and volatilities. Here we assumed a risk-free rate and expected value of the log-normal distribution of 10\% per annum.}
\label{fig-db-put-spread}
\end{figure}

Figure \ref{fig-db-put-spread} shows the resulting diversification benefits for the $10\%$ put spread
as a function of the (upper) strike price given different standard deviations (for the annual log returns).
The shape of the curve results from the competition between the increase in the diversification benefit
from increasing the strike price of the upper put as illustrated in Figure \ref{fig-db-put},
and the loss of diversification from the lower sold put. Note though that these effects are not additive as follows from (\ref{eq-tp}) and the non-linearity of the logarithm.

\subsection{Collar}

The collar is a popular strategy because the downside protection is (partially or fully) financed by selling upside participation (in form of a sold call).
As the collar is the combination of a bought put with a sold call we have (including the underlying)
\be
\Delta=e^{-D\tau} [N(d_1(x,E_p)) -1] - N(d_1(x,E_c)) +1,
\ee
where $E_p$ is the strike of the put, and $E_c$ the strike of the call.
\begin{figure}[htb]
\begin{center}
\includegraphics[width=2.8in]{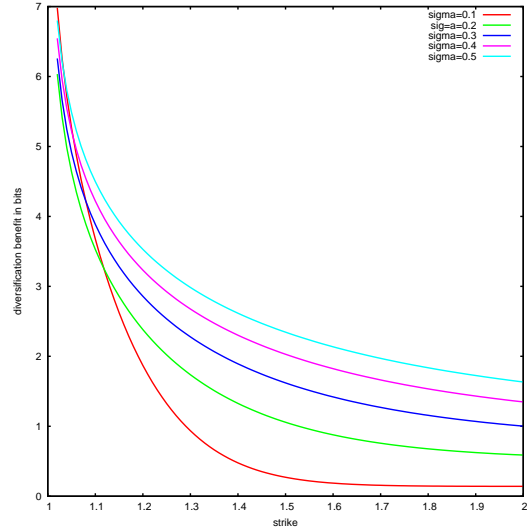}
\end{center}
\caption{Diversification benefit (considered for one year) of a collar position with a two year term. The increase in the diversification
measure is shown for different strikes and volatilities. Here we assumed a risk-free rate and expected value of the log-normal distribution of 10\% per annum.}
\label{fig-db-collar}
\end{figure}

Figure \ref{fig-db-collar} shows the net diversification benefit achieved from a collar in terms of the strike of the upper sold call option. The strike of the
bought put was fixed at 100\% and curves for different volatilities (of the annualized log-return distribution)
are shown.

As the call strike price approaches the put strike price at 100\% all uncertainty is removed (it becomes a synthetic sold future paired with the underlying asset)
and the diversification benefit increases to infinity.
The amount of diversification is notably higher in the case of more volatile assets.

\section{Conclusion}
In this paper we advocate that a measure of portfolio diversification should be a functional of the (subjective) probability distribution for the final portfolio value.
In this way it automatically incorporates asset exposures and asset characteristics.

We propose the negative information entropy of the probability distribution of the final portfolio value as a suitable diversification measure, which
takes all moments into account.

This measure can be applied to all types of distributions and does not require that asset returns or weights are defined.
This means that this measure can deal with holdings which can have positive and negative values, like derivative structures, in a portfolio.

For the simplistic case of independent, normally distributed asset returns, we have demonstrated that the negative information entropy is a logarithmic function of the
portfolio variance. Combining Gaussian assets of equal variance adds an amount to the diversification measure which only depends on the asset weights.

For derivative overlays our result (\ref{eq-tp}) links the future delta to the diversification benefit.

For the more realistic example of log-normal final asset value probability distributions we have given some simple examples illustrating how
derivative-overlays lead to a diversification benefit, as measured by our index.

As protective derivative overlays narrow the probability distribution of the final portfolio value, they increase the diversification as measured by our measure.

We find that for a long-dated put option, the diversification rises rapidly with increasing strike price for less volatile underlying stocks.
For the case of a put spread, this index can be used to select the strike prices that maximizes the net diversification benefit for a given level of volatility.
For the case of a collar, the portfolio diversification benefit achieved increases dramatically as the sold call strike approaches the bought put strike.

\appendix
\section{Understanding Information Entropy}

Let us assume that (discrete) events $i$ occur with likelihood $p_i$ in a repeated experiment. On observing the outcome $i$
the information gained in bits is $-\log_2(p_i)$.

One way to look at this is as follows:
if our state of knowledge is that we judge an object to be equally likely
in one of $2^n$ positions then we gain $n$ bits when we find out where it actually is.

As events occur with probability $p_i$ the average information gain (in bits) is
\be
E = - \sum_i p_i \log_2(p_i),
\ee
which is called the information entropy.

The information entropy has some interesting properties:
\begin{itemize}
\item $E=0$ if $p_j=1$ for one $j$.
\item $E$ takes a maximum value of $\log_2(N)$ if the $N$ possible outcomes are {\em equally} likely.
\item if outcome $j$ is really a combination of different outcomes $j,k$ then
\be
E=E'+p_jE'',
\ee
where $E'$ is the entropy based on $p_j$
(the probability that one of the $k$ events occures)
and $E''$ is the entropy for outcomes $k$, given that $j$ (one of them) occured.
\end{itemize}

Maximum entropy methods generally maximize the information entropy subject to constraints on the probabilities, such as specified moments.

\section{Log-normal assets}
The log-normal distribution with expected value $e^\nu$ is given by the probability distribution
\be
\probdist{A_x}{I} = \frac{1}{x \sqrt{2 \pi \sigma^2}} e^{-\frac{(\ln x -\nu + \sigma^2/2)^2}{2 \sigma^2}},
\ee
where $\sigma^2$ is the variance of the distribution of $\log(x)$.
This yields for the diversification measure
\be
D(A|I)=-\frac 1 2 - \frac 1 2 \log (2 \pi \sigma^2) - \nu + \sigma^2/2.
\ee
It is interesting to note that in contrast to the gaussian case (which depends solely on the variance), the measure here also depends on the expected value.

\section{Effects of Skewness and Kurtosis}
\begin{figure}[htb]
\begin{center}
\includegraphics[width=2.8in]{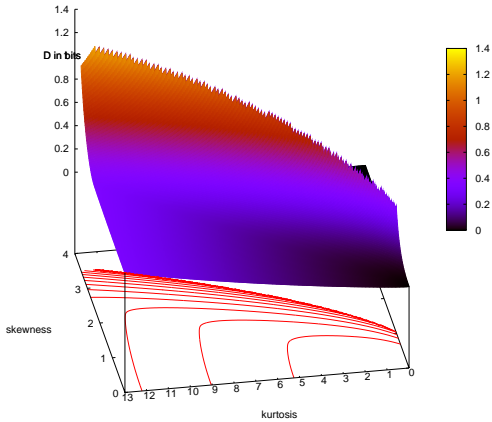}
\end{center}
\caption{Plot showing the diversification index decrease (from the corresponding value of the normal distribution) for maximum entropy
distributions of varying skewness and kurtosis for fixed variance (25\%) (the undefined values have been set to zero.)}
\label{fig-cum-D}
\end{figure}

As the normal distribution is a maximum entropy distribution, any other distribution with equal variance must have more diversification
according to our measure. Figure \ref{fig-cum-D} shows the increase in diversification for maximum entropy distributions of given skewness and kurtosis
for a fixed variance of $\sigma^2=0.25^2$.

\section{Comments on Weight Entropy and Herfindahl Index}
The Weight Entropy and Herfindahl index are both diversification measures solely based on portfolio weights.
We want to investigate the difference between these measures for the two and three asset case. 

For a portfolio with two assets, the Weight Entropy (entropy) and Herfindahl Index (quadratic) measures are shown below in figure \ref{fig-measure-contributions}.  Both are maximal for $w_{1/2}=0.5$.
If the contribution to the overall measure from one asset only is considered, we see (in the lower panel) that large weights effectively give a negative contribution for the quadratic measure, while there
are no negative contributions to the entropy measure.
\begin{figure}[htb]
\begin{center}
\includegraphics[width=3.2in]{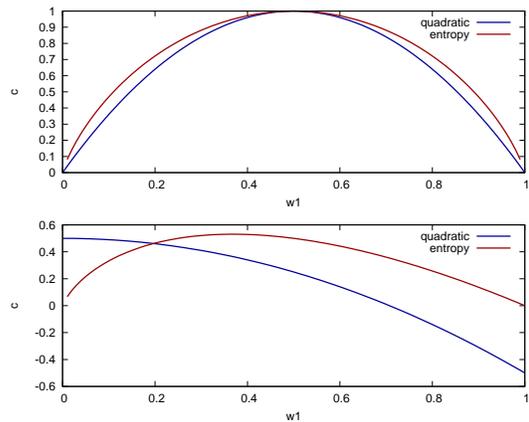}
\end{center}
\caption{ Comparison of entropy and Herfindahl Index ($\sum_{i=1}^N (1/N-w_i^2)$ for two weights $w_1$ and $w_2$ ($w_1+w_2=1$). 
The upper graph shows the overall measure and the lower graph shows the contribution of one weight. The measures have been re-scaled
to cover the same range.}
\label{fig-measure-contributions}
\end{figure}

\begin{figure}[htb]
\begin{center}
\includegraphics[width=3.2in]{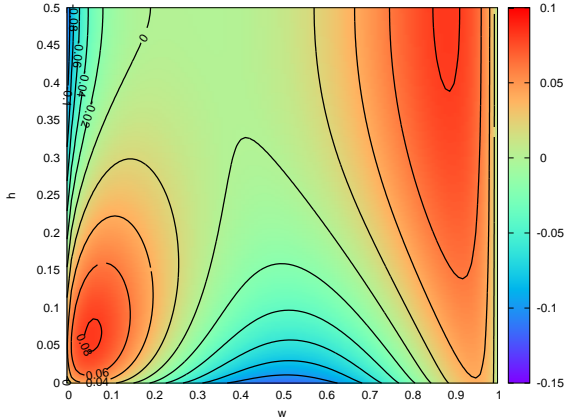}
\end{center}
\caption{ Comparison of entropy $-\sum_{i=1}^3 w_i \log(w_i)/\log(3)$
and re-scaled Herfindahl Index ($\frac{3}{2}\sum_{i=1}^3 (1/N-w_i^2)$) for three weights. Here $w_2=h(1-w_1)$, which defines $h$ in terms of the weights.
Both measures coincide with their maximum at $w_1=1/3$ and $h=0.5$.
Shown is the difference of both moeasures --- red indicates a higher entropy measure of diversification relative to the quadratic measure. Values for $h'>0.5$ are equal to values for $h=1-h'$. 
}
\label{fig-three-asset-contour}
\end{figure}

Figure \ref{fig-three-asset-contour} shows the differences for the three asset case. There are large areas where
the quadratic measure (Herfindahl Index) understates diversification by up to $10\%$ of the maximum value compared to the entropy measure.
On the other hand, around $w_1=0.5$ and $w_2=h(1-w_1)=0$ (and equivalently for $w_1=0$ and $w_2=0.5$) the quadratic measure
overstates diversification relative to the entropy measure.

Both measures can be extended to the correlated gaussian case by considering the exposure to the ``eigen-vector portfolios'' of the correlation matrix instead of the
actual asset weights. This, however, still does not account for varying risk profiles of the assets (or eigen-vector portfolios).

\end{document}